\documentclass[aps,amssymb,pra,showpacs,twocolumn]{revtex4}

\usepackage{epsf}
\usepackage{amsmath}
\usepackage{graphicx}

\begin{document}

\title{Low-energy fusion caused by an interference}

\author{B. Ivlev} 

\affiliation
{Universidad Aut\'onoma de San Luis Potos\'{\i}, San Luis Potosi, Mexico}


\begin{abstract}
Fusion of two deuterons of room temperature energy is studied. The nuclei are in vacuum with no connection to any external source (electric or magnetic field, illumination, surrounding matter, traps, etc.) which may accelerate them. The energy of the two nuclei is conserved and remains small during the motion through the Coulomb barrier. The penetration through this barrier, which is the main obstacle for low-energy fusion, strongly depends on a form of the incident flux on the Coulomb center at large distances from it. In contrast to the usual scattering, the incident wave is not a single plane wave but the certain superposition of plane waves of the same energy and various directions, for example, a convergent conical wave. As a result of interference, the wave function close to the Coulomb center is determined by a cusp caustic which is probed by de Broglie waves. The particle flux gets away from the cusp and moves to the Coulomb center providing a not small probability of fusion (cusp driven tunneling). Getting away from a caustic cusp also occurs in optics and acoustics.

\end{abstract} \vskip 1.0cm

\pacs{25.45.-z, 03.65.Xp, 03.65.Sq} 

\maketitle

\section{INTRODUCTION}
\label{intr}
The aspects of nuclear fusion are discussed, for instance, in Refs.~\cite{HERM,TAUB,STORMS1,HUKE1,PUTT,WID,HUKE2,SHELD,KIM,STORMS2} and references therein. Here we outline principal phenomena associated with nuclear fusion. The main difficulty is getting the nuclei close enough to fuse since they should overcome a high Coulomb barrier.

There are two ways to pass the Coulomb barrier, to accelerate the nuclei up to a high energy comparable with the barrier height (of the order of $1~{\rm MeV}$) or to pass the barrier via quantum tunneling. When the energy is not high, the probability of tunneling of the nuclei through the Coulomb barrier is extremely small according to the theory of Wentzel, Kramers, and Brillouin (WKB) \cite{LANDAU1}. So only high energy nuclei can fuse.

Therefore, high energy nuclei is the leading idea of fusion technique. We mention an acceleration of initially cold deuterons by a strong electric field using a pyroelectric crystal \cite{PUTT}.

It is surprising to claim that {\it two bare deuterons}, that is isolated from everything, {\it of room temperature energy} are able, in principle, to penetrate the Coulomb barrier with a not small probability and to subsequently fuse. This statement is counterintuitive. A tennis ball cannot penetrate through a brick wall. This correlates with the usual underbarrier physics led by the philosophy of addition of probabilities but not amplitudes. The proposed phenomenon of barrier penetration is based ultimately on interference that is on addition of amplitudes.

We really deal with two low-energy nuclei in vacuum. There are no external sources (electric or magnetic field, illumination, surrounding matter, traps, etc.) which may accelerate them. The energy of the two nuclei, in the system of center of mass, is conserved and remains small during the motion through the Coulomb barrier. This is a substantial difference from usual schemes to push in action a mechanism of low-energy fusion by some local heating or acceleration of nuclei.

The conventional scattering problem is a study of reflection of the incident flux which is a plane wave coming from large distances \cite{LANDAU1}. In the Coulomb field scattering relates to the Rutherford formula and the wave function is exponentially small at the center.

What happens when the incident flux at large distances is not just a single plane wave?

The wave function close to the Coulomb center can be substantially modified when the incident flux is a superposition of plane waves with the same energy and various directions, for example, a convergent conical wave. In this case classical trajectories are reflected, due to Coulomb forces, from the certain surface surrounding the cone axis. This surface is called caustic \cite{LANDAU3} and it is terminated by the cusp directed to the center. Classical trajectories fill out the space restricted by the caustic surface. Each incident trajectory reflects from the caustic surface, continues inside, and pierces its opposite part. So the space outside the caustic is not an absolute shadow from the classical standpoint.

Classical trajectories, which pierce the surface of the cusp caustic, do not reach the Coulomb center since they are reflected from another caustic closer to the center. So the Coulomb center is in the absolute shadow. Otherwise it would be strange to get the center reachable within classical physics.

Unlike classical physics, a caustic shadow is not completely "empty`` due to an exponential decay of the wave function outside the caustic surface \cite{LANDAU3}. The proper outer flux of the cusp caustic is directed along it. When the flux reaches the cusp it gets away from it and moves towards the center. It pierces the caustic which reflects the above classical flux propagated after piercing the cusp caustic. Then it reaches the Coulomb center providing a not exponentially small probability of tunneling. The cusp driven flux does not exist in classical physics. The analogous flux, getting away from a caustic cusp, is also formed in optics and acoustics.

In this paper the semiclassical method, based on Hamilton-Jacobi equation \cite{LANDAU1,LANDAU2}, was used. The phenomenon of interference in tunneling was already proposed in Refs.~\cite{IVLEV1,IVLEV6,IVLEV7,IVLEV5}. Multi-dimensional tunneling was studied, in particular, in Refs.~\cite{COLEMAN1,COLEMAN2,SCHMID1,SCHMID2,SAKITA,LEGGETT,HELLER,DYK,ANKER}. Influence of interference on tunneling through nonstationary barriers was studied in Refs.~\cite{IVLEV2,IVLEV4}. The role of interference in $\alpha$ decay was studied in Ref.~\cite{IVLEV3}.

Experimental conditions to form the required incident flux are briefly discussed in Sec.~\ref{disc}.
\section{GENERAL VIEW}
\label{gen}
Two examples of nuclear fusion are below. The deuterium-deuterium fusion occurs by the scheme
\begin{equation}
\label{1}
^{2}_{1}H+\,^{2}_{1}H\rightarrow \,^{3}_{2}He+n+3.27\,{\rm Mev}.
\end{equation}
The deuterium-tritium fusion releases more energy
\begin{equation}
\label{2}
^{2}_{1}H+\,^{3}_{1}H\rightarrow \,^{4}_{2}He+n+17.59\,{\rm Mev}.
\end{equation}
The problem can be separated by two steps (i) overcoming the Coulomb barrier by two nuclei with masses $M_{1}$ and $M_{2}$ and (ii) a subsequent nuclear fusion at the short distance $R_{0}\sim 10^{-13}\,{\rm cm}$. The former is a motion of a particle with the reduced mass $M=M_{1}M_{2}/(M_{1}+M_{2})$ described by the Schr\"{o}dinger equation
\begin{equation}
\label{3}
-\frac{\hbar^2}{2M}\,\frac{\partial^{2}\psi}{\partial\vec R^{2}}+\frac{e^2}{R}\,\psi=E\psi,
\end{equation}
where $R$ is the inter-nuclear distance \cite{LANDAU1}. Below we measure length in the units of $2e^2/E$. The wave function can be written in the form
\begin{equation}
\label{4}
\psi=\exp(iB\sigma),
\end{equation}
where $\sigma$ satisfies the equation
\begin{equation}
\label{6}
\frac{1}{4}\,\left(\frac{\partial\sigma}{\partial\vec R}\right)^2+\frac{1}{2R}-\frac{i}{4B}\,\frac{\partial^{2}\sigma}{\partial\vec R^{2}}=1.
\end{equation}
The parameter
\begin{equation}
\label{5}
B=\frac{e^2}{\hbar c}\,\sqrt{\frac{2Mc^2}{E}}
\end{equation}
is supposed to be large which corresponds to semiclassical approximation. In this case one can ignore the last term in Eq.~(\ref{6}) and it goes over into the equation of Hamilton-Jacobi when $\sigma$ is the classical action divided by $\hbar B$ \cite{LANDAU1,LANDAU2}.

In a spherically symmetric case the first term in Eq.~(\ref{6}) is just $(\partial\sigma/\partial R)^{2}/4$. The wave function exponentially decays inside the Coulomb barrier. Accordingly, the
imaginary part of $\sigma$, related to a motion under the barrier, is
\begin{equation}
\label{7}
\sigma=i\sqrt{2}\int^{1/2}_{R_{0}}dR\,\sqrt{\frac{1}{R}-2}\simeq\frac{i\pi}{2},
\end{equation}
where we ignore the distance $R_{0}$ which is short compared to scales of the Coulomb motion. Eq.~(\ref{7}) follows from the WKB approximation The probability of fusion is one to overcome the Coulomb barrier
\begin{equation}
\label{8}
w=\Bigg|\frac{\psi(R_{0})}{\psi(\infty)}\Bigg|^{2}\sim\exp(-\pi B).
\end{equation}

One can estimate the fusion probability (\ref{8}), for example, for the reaction (\ref{1}). In this case the reduced mass is $M=M_{D}/2$ where the deuteron mass is defined as
$M_{D}c^2\simeq 1.87\times 10^{9}\,{\rm eV}$. The nuclear Bohr radius is ${\hbar^{2}/M}e^2\simeq 2.88\times 10^{-12}{\rm cm}$. At the energy of two deuterons $E=T$, where $T=300\,{\rm K}\simeq 2.58\times 10^{-2}\,{\rm eV}$ relates to the room temperature, the de Broglie wave length is $\lambda=2\pi\hbar/\sqrt{2M_DE}\simeq 1.26\,\AA$. Two deuterons enter the Coulomb barrier at the distance $e^2/E\simeq 557\,\AA$ between them. With these parameters one can estimate $\pi B\simeq 6174.9$. This corresponds to the probability $w\sim 10^{-2682}$ which is in accordance with usual estimates for low-energy fusion.
\section{SCATTERING BY THE COULOMB CENTER}
\label{scat}
The above estimate of nonphysically small probability of low-energy fusion corresponds to the usual WKB approximation. For a spherically symmetric incident flux the problem is generic with
one-dimensional one since radial and spherical parts of the wave function are separated.

There is another remarkable case of variables separation in the Coulomb field. It occurs in parabolic coordinates
\begin{equation}
\label{9}
\xi=\sqrt{r^2+z^2}+z,\hspace{0.5cm}\eta=\sqrt{r^2+z^2}-z.
\end{equation}
In this case the total wave function has the form
\begin{equation}
\label{10}
\psi(\xi,\eta)=f(\xi)\varphi(\eta)\exp(im\varPhi),
\end{equation}
where $\varPhi$ is the azimuthal angle and $m$ is the magnetic quantum number. The use of spherical coordinates is less convenient for our purposes due to the discrete summation on azimuthal quantum number $l$ as in the problem of Regge poles \cite{REG}. Below we consider an axially symmetric case when $m=0$ and the Schr\"{o}dinger equation has the form
\begin{equation}
\label{10a}
-\frac{1}{4B^{2}r}\,\frac{\partial}{\partial r}\left(r\frac{\partial\psi}{\partial r}\right)-\frac{1}{4B^2}\,\frac{\partial^2\psi}{\partial z^2}+\frac{\psi}{2\sqrt{r^2+z^2}}=\psi.
\end{equation}
The parts $f$ and $\varphi$ obey the equations
\cite{LANDAU1}
\begin{eqnarray}
\label{11a}
-\frac{1}{B^2\xi}\,\frac{\partial}{\partial\xi}\left(\xi\frac{\partial f}{\partial\xi}\right)+\frac{1+\beta}{\xi}\,f=f\\
\label{11b}
-\frac{1}{B^2\eta}\,\frac{\partial}{\partial\eta}\left(\eta\frac{\partial\varphi}{\partial\eta}\right)-\frac{\beta}{\eta}\,\varphi=\varphi\,,
\end{eqnarray}
where $\beta$ is a constant connected with the variable separation.
\begin{figure}
\includegraphics[width=7.0cm]{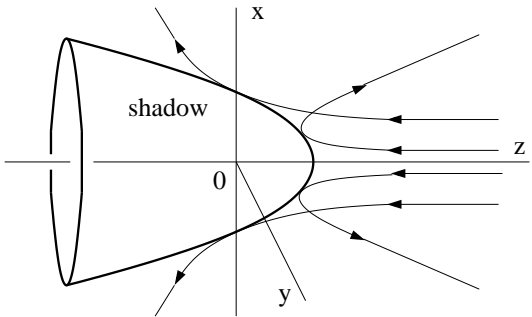}
\caption{\label{fig1}The flux on the Coulomb center (zero coordinates) is reflected from the caustic surface defined by Eq.~(\ref{15}). Four trajectories from the infinite set are shown.}
\end{figure}

Each of Eqs.~(\ref{11a}) and (\ref{11b}) have two independent solutions. When $\xi$ is not small ($r$ is not small at $z<0$) and $\eta$ is not small ($r$ is not small at $0<z$) one can use the
semiclassical approach for the wave functions which is not valid at very small $r$ only \cite{LANDAU1}. In this limit two branches of $f$ have the asymptotic form
\begin{equation}
\label{12}
f_{\mp}(\xi,\beta)\sim\exp\left(\mp iB\int^{\xi}_{1+\beta}d\xi_{1}\sqrt{1-\frac{1+\beta}{\xi_{1}}}\,\,\right)
\end{equation}
and analogously two branches of $\varphi$ are
\begin{equation}
\label{13}
\varphi_{\mp}(\eta,\beta)\sim\exp\left(\mp iB\int^{\eta}_{0}d\eta_{1}\sqrt{1+\frac{\beta}{\eta_{1}}}\,\,\right).
\end{equation}
Withing the semiclassical accuracy we do not specify preexponential factors.

The velocity field can be studied by Newtonian trajectories. According to classical mechanics \cite{LANDAU2}, velocities in the $\xi$ and $\eta$ direction are proportional to $\sqrt{1-(1+\beta)/\xi}$ and $\sqrt{1+\beta/\eta}$ respectively. This sets a velocity field in the plane $\{\xi,\eta\}$. Each point in the plane $\{\xi,\eta\}$ belongs to one or a few
classical trajectories $\eta(\xi)$.

In the usual scattering problem the incident flux is solely a plane wave from large positive $z$. This situation corresponds to $\beta=-i/B$ ($\beta\simeq 0$ in the semiclassical approximation) when the exact solution of Eq.~(\ref{11b}) is $\varphi_{+}=\exp(iB\eta)$ \cite{LANDAU1}. The combination $f_{-}(\xi)\varphi_{+}(\eta)$ results in the incident plane wave $\psi\sim\exp(-2iBz)$ at large positive $z$. Analogously, the combination $f_{+}(\xi)\varphi_{+}(\eta)$ leads to the scattered wave $\psi\sim\exp(2iB\sqrt{r^2+z^2}\,)$ far from the center.

Classical trajectories are shown in Fig.~\ref{fig1}. They are reflected from the caustic surface where the velocity $\partial\xi/\partial t=0$ \cite {LANDAU3}. See also \cite{NELS}. This happens at $\xi=1+\beta\simeq 1$. The caustic surface in Fig.~\ref{fig1} is axially symmetric, since $m=0$, and, as follows from Eq.~(\ref{9}), is given by
\begin{equation}
\label{15}
2z=1-r^{2}.
\end{equation}
Along the caustic the normal momentum is zero. The tangent momentum is real and determined by $\sqrt{1+\beta/\eta}\simeq 1$. The wave function exponentially decays inside shadow region in Fig.~\ref{fig1}.
\section{FORMULATION OF THE PROBLEM}
\label{form}
Below we explore the superposition of functions (\ref{12}) and (\ref{13}) with different $\beta$
\begin{equation}
\label{16}
\psi_{I}(\xi,\eta)=\int d\beta\,c_{I}(\beta)f_{\mp}(\xi,\beta)\left[\varphi_{-}(\eta,\beta)+\varphi_{+}(\eta,\beta)\right]
\end{equation}
\begin{eqnarray}
\nonumber
&&\psi_{II}(\xi,\eta)=\int d\beta\,c_{II}(\beta)f_{\mp}(\eta,-1-\beta)[\varphi_{+}(\xi,-1-\beta)\\
&&+\varphi_{-}(\xi,-1-\beta)]
\label{16a}
\end{eqnarray}
with certain weight functions $c_{I,II}(\beta)$ specified further. The total wave function is
\begin{equation}
\label{16b}
\psi(\xi,\eta)=\psi_{I}(\xi,\eta)+\psi_{II}(\xi,\eta).
\end{equation}

As shown below, the function $\psi_{I}$ is exponentially small at $z<0$ and the function $\psi_{II}$ is exponentially small at $0<z$. The functions $\varphi_{\mp}(\eta)$ contain the singular 
part $\ln\eta$ at small argument. The same is valid for functions $f_{\mp}(\xi)$ \cite{LANDAU1}. The functions $\varphi_{\mp}(\eta)$ are chosen in the way that the non-physical singularity is cancelled in the combination $\varphi_{-}(\eta)+\varphi_{+}(\eta)$ which is finite at small $\eta$. It is not difficult to show that at large $\eta$ the preexponential factor of $\varphi_{-}$
is $C/\sqrt{\eta}$ and the preexponential factor of $\varphi_{+}$ is $iC/\sqrt{\eta}$ where $C$ is a constant. According to Eqs.~(\ref{16}) and (\ref{16a}), the main parts of the total wave function (\ref{16b}) are not singular close to $r=0$ at all $z$. Singularities of the exponentially small functions ($\psi_{I}$ at $z<0$ and $\psi_{II}$ at $0<z$) are not important since they are compensated by exponentially small corrections of the main parts.

The interference, described by Eqs.~(\ref{16}) and (\ref{16a}), of partial waves with different $\beta$ strongly depends on forms of weight functions $c_{I,II}(\beta)$. In particular, these functions define an incident flux far from the Coulomb center. Details of this flux, and therefore $c_{I,II}(\beta)$, are set by experimental conditions. Below we use the forms
\begin{equation}
\label{17}
c_{I,II}(\beta)\sim\exp\left(\mp iB\alpha\beta\right),
\end{equation}
where $\alpha$ is a fixed positive parameter. The related incident flux is discussed in Sec.~\ref{far}.

According to Eqs.~(\ref{12}) and (\ref{13}), there are four ($i=1,2,3,4$) semiclassical branches of the wave function $\psi_{I}$ (analogously, of $\psi_{II}$)
\begin{equation}
\label{18}
\psi_i=\int a_{i}\exp(iB\sigma_i)d\beta,
\end{equation}
where the preexponential factors $a_{i}$ are not important for the semiclassical approach used. $\sigma_i$ are defined by the following
\begin{equation}
\label{19}
\sigma_{2,3}=\mp\int^{\xi}_{1+\beta}\sqrt{1-\frac{1+\beta}{\xi_1}}\,d\xi_1+\int^{\eta}_{0}\sqrt{1+\frac{\beta}{\eta_1}}\,d\eta_1-i\alpha\beta
\end{equation}
and $\sigma_{1,4}+i\alpha\beta=-(\sigma_{3,2}+i\alpha\beta)$.

The large parameter $B$ provides validity of semiclassical approximation and the $\beta$-integration can be done by the saddle method. This means that for each $\xi$ and $\eta$ one can determine
the certain $\beta(\xi,\eta)$ from the condition $\partial\sigma_i/\partial\beta=0$. For $\sigma_1$ this condition reads
\begin{equation}
\label{20}
\sqrt{\frac{\xi}{1+\beta}}+\sqrt{\frac{\xi}{1+\beta}-1}=\left(\sqrt{\frac{\eta}{\beta}+1}+\sqrt{\frac{\eta}{\beta}}\,\right)\exp({\alpha}).
\end{equation}
The analogous condition for $\sigma_2$ differs from (\ref{20}) by the sign at $\sqrt{\eta/\beta}$. The condition for $\sigma_3$ differs from (\ref{20}) by the sign of $\alpha$. The condition for $\sigma_4$ is obtained if to make the both changes in Eq.~(\ref{20}).

The above semiclassical solutions also follow from the equation of Hamilton-Jacobi. Eq.~(\ref{6}) goes over into this equation if to drop the second derivative. The Hamilton-Jacobi equation in parabolic coordinates
\begin{equation}
\label{21}
\frac{\xi}{\xi+\eta}\left(\frac{\partial\sigma}{\partial\xi}\right)^{2}+\frac{\eta}{\xi+\eta}\left(\frac{\partial\sigma}{\partial\eta}\right)^{2}+\frac{1}{\xi+\eta}=1
\end{equation}
allows separation of variables used above. Equations (\ref{19}) for $\sigma_i$ with the conditions (\ref{20}) are equivalent to a general integral of the Hamilton-Jacobi equation (\ref{21})
\cite{LANDAU2}.

To obtain derivatives, $\partial\sigma_i/\partial\xi$ and $\partial\sigma_i/\partial\eta$ one should differentiate in Eq.~(\ref{19}) with respect to the upper integration limits only. For $\sigma _1$, for example, we have
\begin{eqnarray}
\label{21a}
&&\frac{\partial\sigma_{1}(\xi,\eta)}{\partial\xi}=-\sqrt{1-\frac{1+\beta(\xi,\eta)}{\xi}}\\
\label{21b}
&&\frac{\partial\sigma_{1}(\xi,\eta)}{\partial\eta}=-\sqrt{1+\frac{\beta(\xi,\eta)}{\eta}}\,.
\end{eqnarray}
The Cartesian derivatives
\begin{equation}
\label{22}
\frac{\partial\sigma_i}{\partial z}=\frac{2}{\xi+\eta}\left(\xi\frac{\partial\sigma_i}{\partial\xi}-\eta\frac{\partial\sigma_i}{\partial\eta}\right)
\end{equation}
and
\begin{equation}
\label{23}
\frac{\partial\sigma_i}{\partial r}=\frac{2\sqrt{\xi\eta}}{\xi+\eta}\left(\frac{\partial\sigma_i}{\partial\xi}+\frac{\partial\sigma_i}{\partial\eta}\right)
\end{equation}
are proportional to classical velocities.

Now one can outline the problem. From Eq.~(\ref{20}) and its analogs one has to determine $\beta(\xi,\eta)$ and to insert it into Eq.~(\ref{19}) and its analogs. It will be a semiclassical solution to be found.
\begin{figure}
\includegraphics[width=8.3cm]{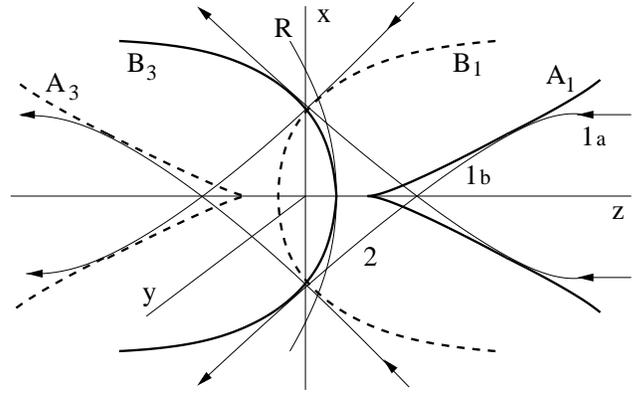}
\caption{\label{fig2}Distribution of the particle flux associated with the wave function $\psi_{I}$. Caustic surfaces are axially symmetric with respect to the $z$ axis. The curves represent intersection of caustics and the $\{x,z\}$ plane. The caustic $A_{1}$ touches the $z$ axis at the point $z_{s}$. The caustic $B_3$ intersects the $z$ axis at the point $z=1/2$. On the dashed caustics $\psi_{I}$ is exponentially small. On the curve $R$ the branch $\psi_{2}$ is converted into $\psi_{3}$.}
\end{figure}
\section{CAUSTICS}
\label{caust}
A condition of applicability of semiclassical approximation is well known. A wave length should be a smooth function of coordinates \cite{LANDAU1}. In other words, quantum corrections to the Hamilton-Jacobi equation (\ref{21}), which contain $(1/B)\partial^{2}\sigma/\partial\xi^{2}$ and $(1/B)\partial^{2}\sigma/\partial\eta^{2}$, should be small. There are various situations when those derivative are not small. First, it happens near a classical turning point where classical momenta ($\sim\partial\sigma/\partial\xi$) are proportional to square root of the distance to this point.

In addition to that, the classical momentum can be not zero but nevertheless its spatial derivative turns to infinity. As follows from Eqs.~(\ref{21a}) and (\ref{21b}), this may happen when
$\partial\beta/\partial\xi$ or $\partial\beta/\partial\eta$ become large. This condition specifies a certain surface, called caustic, in the three-dimensional space where classical trajectories are tangent to it \cite{LANDAU3}. In our case caustic surfaces have axial symmetry with respect to the $z$ axis and one can study just caustic curves in the $\{r,z\}$ plane. The caustic condition can be obtained from Eq.~(\ref{20}) by the formal condition $\partial\xi/\partial\beta=0$ at a fixed $\eta$. After a little algebra the caustic condition reads
\begin{eqnarray}
\label{24}
&&\frac{2\xi}{1+\beta}=2(1+\beta)\cosh^{2}\alpha-\sinh^{2}\alpha\\
\nonumber
&&\pm\sinh\alpha\sqrt{\sinh^{2}\alpha+4\beta(1+\beta)\cosh^{2}\alpha}\,,\\
\nonumber
&&\frac{2\eta}{\beta}=2\beta\cosh^{2}\alpha+\sinh^{2}\alpha\\
\nonumber
&&\mp\sinh\alpha\sqrt{\sinh^{2}\alpha+4\beta(1+\beta)\cosh^{2}\alpha}\,,
\end{eqnarray}
which determines the caustic form $\eta(\xi)$ if to exclude $\beta$ in these equations.

The upper sign in Eq.~(\ref{24}) relates to the branch $\sigma_1$ and produces the caustics $A_1$ and $B_1$ in Fig.~\ref{fig2}. The lower sign in Eq.~(\ref{24}) relates to the branch $\sigma_3$ and results in the caustics $A_3$ and $B_3$ in Fig.~\ref{fig2}.

The caustic $A_1$ at small $r$ corresponds to small $\beta$ and to large $\beta$ at large $r$. One can easily obtain the shape of the caustic $A_1$ in limiting cases
\begin{equation}
\label{25}
r=\frac{4\sqrt{2}}{\sinh 2\alpha}\left(\frac{z-z_s}{3}\right)^{3/2},\hspace{0.4cm}(z-z_s)\ll z_s,
\end{equation}
\begin{equation}
\label{26}
r=\frac{z}{\sinh\alpha}\left(1-\sqrt{\frac{2z_s}{z\tanh\alpha}}\,\right),\hspace{0.4cm}z_s\ll z,
\end{equation}
where $2z_s=\cosh^{2}\alpha$. The caustic $B_1$ corresponds to $\beta<-1$. At small $r$ it should be $(-1-\beta)\ll 1$ and at large $r$ the parameter $(-\beta)$ is large. The form of the caustic $B_1$ in limiting cases is
\begin{equation}
\label{27}
r=\sinh\alpha\sqrt{1+2z},\hspace{0.4cm}(2z+1)\ll 1,
\end{equation}
\begin{equation}
\label{28}
r=\frac{z}{\sinh\alpha}\left(1+\sqrt{\frac{2z_s}{z\tanh\alpha}}\,\right),\hspace{0.4cm}1\ll z.
\end{equation}
The pair of caustics $\{A_3,B_3\}$ is a mirror reflection of the pair $\{A_1,B_1\}$ with respect to the $x$ axis as shown in Fig.~\ref{fig2}. The caustics $A_1$ and $A_3$ have a cusp shape close to the points $z=\pm z_{s}$.

Along caustics ${\rm Im}\,\sigma={\rm const}$ and the momentum is tangent to the caustic, that is, along each caustic,
\begin{equation}
\label{29}
\left(\frac{\partial{\rm Re}\sigma}{\partial r}\right)=\left(\frac{\partial{\rm Re}\sigma}{\partial z}\right)\frac{\partial r}{\partial z}\,.
\end{equation}
\section{DISTRIBUTION OF THE FLUX $\psi_{I}$}
\label{distr}
In this section we specify the velocity field associated with the wave function $\psi_{I}$. First, we focus on the region $(z-z_{s}),r\ll 1$ where the caustic $A_1$ in Fig.~\ref{fig2} is about to touch the $z$ axis. In this case the parameter $\beta$ is small and Eq.~(\ref{20}) can be written in the form
\begin{equation}
\label{30}
\beta^{3/2}_{1}+\frac{z_{s}-z}{z_s}\sqrt{\beta_1}+r\sqrt{\frac{2}{z_s}}\tanh\alpha=0,
\end{equation}
which is cubic with respect to $\sqrt{\beta_1}$. We ascribe the index ``1'' to $\beta$ to emphasize its connection to the branch $\sigma_1$.

Two physical solutions of Eq.~(\ref{30}) at $(z-z_s)\ll 1$ are
\begin{equation}
\label{31}
\beta_{1a,b}=\frac{z-z_s}{3z_s}\left[1\mp\sqrt{\frac{1}{3}-\left(\frac{3r\sinh 2\alpha}{4\sqrt{2}}\right)^{2}\left(\frac{1}{z-z_s}\right)^{3}}\right]
\end{equation}
which are valid close to the caustic $A_1$, that is, when the square root in Eq.~(\ref{31}) is small. Two signs of the square root provide two opposite velocities normal to the caustic $A_1$ from the classical side. Close to the $z$ axis
\begin{equation}
\label{32}
\beta_{1a}=\left(\frac{r\sinh\alpha}{z-z_s}\right)^{2},\hspace{0.2cm}\beta_{1b}=\frac{z-z_s}{z_s},\hspace{0.2cm}r^2\ll(z-z_s)^3\ll 1
\end{equation}
Now it easily follows from Eq.~(\ref{20}) that
\begin{eqnarray}
\label{33}
&&\beta_{1a}=\frac{r^2}{8z}\left(\frac{\sqrt{2z}\sinh\alpha+\sqrt{2z-1}\cosh\alpha}{z-z_s}\right)^{2},\\
\nonumber
&&\beta_{1b}=\frac{z-z_s}{z_s}-\frac{rz\sqrt{2}\tanh\alpha}{z_s\sqrt{z-z_s}},\hspace{0.4cm}r^2\ll(z-z_s)^3.
\end{eqnarray}
As one can see, Eqs.~(\ref{33}) go over into Eqs.~(\ref{32}) when $z$ is close to $z_s$.
\subsection{Incident flux}
Now one can analyze what happens to the incident flux (from the right in Fig.~\ref{fig2}). It is described by the action $\sigma_{1a}$ where the index $a$ means that one has to substitute $\beta_{1a}$ in Eq.~(\ref{19}) for $\sigma_{1}$. As follows from Eqs.~(\ref{20}), (\ref{22}), and (\ref{23}), at $z$ not too close to $z_s$
\begin{eqnarray}
\label{34}
&&\frac{\partial\sigma_{1a}}{\partial z}=-2\sqrt{1-\frac{1}{2z}}\,,\hspace{0.4cm}r\ll 1,\hspace{0.4cm}z_s<z\\
\nonumber
&&\frac{\partial\sigma_{1a}}{\partial r}=-\frac{r}{z}\Bigg[\sqrt{1-\frac{1}{2z}}\\
\label{35}
&&+\sqrt{1+\left(\frac{\sqrt{2z}\sinh\alpha+\sqrt{1-2z}\cosh\alpha}{2(z-z_s)}\right)^{2}}\,\,\Bigg].
\end{eqnarray}

It is instructive to consider $\sigma_{1}$ on a classical trajectory which is just a curve in the entire space. The trajectory $1a$, associated with the branch $\sigma_{1a}$, is shown in Fig.~\ref{fig2}. Trajectories of the type $1a$ do not intersect the $z$ axis since, according to Eq.~(\ref{35}), the normal to this axis momentum becomes zero on it.
\subsection{Flux associated with the caustic $A_1$}
The trajectory $1a$ is reflected from the caustic $A_1$ and goes over after the reflection into the trajectory $1b$ as shown in Fig.~\ref{fig2}. The related branch is $\sigma_{1b}$, that is $\sigma_1$ with the above $\beta_{1b}$. As follows from Eqs.~(\ref{20}), (\ref{22}), and (\ref{23}), at $z_{s}<z$
\begin{eqnarray}
\nonumber
&&\frac{\partial\sigma_{1b}}{\partial z}=-2\tanh\alpha-\frac{r}{z}\,\sqrt{\frac{z_s}{2z(z-z_s)}}\,\hspace{0.3cm}r\ll 1,\\
\label{36}
&&\frac{\partial\sigma_{1b}}{\partial r}=-\sqrt{\frac{2(z-z_s)}{zz_s}}+\frac{rz_s\tanh\alpha}{z(z-z_s)}\,.
\end{eqnarray}
According to Eq.~(\ref{36}), the trajectory $1b$ intersects the $z$ axis since the momentum, normal to this axis, is finite.

After intersection of the $z$ axis, the trajectory $1b$ turns to the trajectory 2 marked in Fig.~\ref{fig2}. The related $\partial\sigma_{2}/\partial z$ and $\partial\sigma_{2}/\partial r$ differ from Eq.~(\ref{36}) by the sign of square roots. This description of the branch 2 is valid for $r\ll 1$ and $z_s<z$. At $1/2<z<z_s$ and $r\ll 1$, $\partial\sigma_{2}/\partial z$ has the same form (\ref{34}) but $\partial\sigma_{2}/\partial r$ differs from (\ref{35}) by the sign of the second square root. Since $\partial\sigma_{2}/\partial r$ is zero at $r=0$, the part $\sigma_{2}$ behaves as one-dimensional one on the $z$ axis. Namely, it is related to the classical turning point $z=1/2$ where the caustic $B_{3}$ intersects the $z$ axis.

At $z_s>z\sim 1$ equations can be formally obained from Eq.~(\ref{36})
\begin{eqnarray}
\nonumber
&&\frac{\partial\sigma_{1}}{\partial z}=-2\tanh\alpha-\frac{ir}{z}\,\sqrt{\frac{z_s}{2z(z_s-z)}}\,,\hspace{0.2cm}r\ll 1\\
\label{37}
&&\frac{\partial\sigma_{1}}{\partial r}=i\sqrt{\frac{2(z_s-z)}{zz_s}}-\frac{rz_s\tanh\alpha}{z(z_s-z)}\,.
\end{eqnarray}
Outside the caustic $A_{1}$ one can drop down the index $b$. At small $r$ and $z$ close to the cusp point $z_{s}$ the action (\ref{37}) can be written in the form
\begin{equation}
\label{37a}
\sigma_{1}=-2z\tanh\alpha+\frac{ir}{z_{s}}\sqrt{2(z_{s}-z)}.
\end{equation}
We see that the branch $\psi_{1}$ decays inside the shadow of the caustic $A_{1}$ and also decays away from the $z$ axis. Eqs.~(\ref{37}) and (\ref{37a})   are not valid at very small $r$ (roughly $r\sim 1/B$) when the second term in (\ref{37a}) does result in a large correction to the action $B\sigma_{1}$. At those small $r$ the branch $\psi_{1}$ has the singularity $\ln r$, mentioned in Sec.~\ref{form}, which has to be compensated by another branch having the same singularity but increasing from the $z$ axis. This branch is absent in Eqs.~(\ref{37}) and (\ref{37a}). What happens to this branch?

Values of the parameter $\beta$ used above, correspond to a saddle point when $\partial\sigma/\partial\beta=0$. The second derivative determines properties of the saddle in the complex plane of $\beta$. At $r=0$ the inclination (with respect to the axis of real $\beta$) of the steepest descent is $(-\pi/4)$. The $\beta$-integration can be chosen along the imaginary direction from the saddle point where the action is real. At large $\beta$, as follows from Eq.~(\ref{19}), the action for the branch 1 is real (independently of $r$) along the line determined by the condition ${\rm Re}\beta(\pi/2-i\alpha)=0$. So at $r=0$ the $\beta$-integration can be taken along the total path where the action is real. This path contains the saddle resulting in a wave function which is not exponentially small.

At small $r$ one can slightly deform (compared to the limit of $r=0$) the contour of the $\beta$-integration to get again a real action on it. But in this case the saddle is outside the contour. For the branch (\ref{37}), decaying from the $z$ axis, the wave function at the saddle point is exponentially small compared to the contour with a real action. Therefore the $\beta$-integration along the steepest descent reduces the branch. On the other hand, the branch, increasing with $r$, relates to the saddle point where the wave function is exponentially large compared to its values on the contour of the real action. Therefore the entire integration path is on the same descent side of the saddle. In this case the integration is not reduced to the steepest descent from the saddle point and results in a strong reduction of the branch. This does not happen at very small $r$ where saddle points of two branches do not differ within the semiclassical accuracy and the logarithms are compensated.

So Eqs.~(\ref{37}) and (\ref{37a}), valid at $1/B<r$, describe the correct wave function which is localized near the $z$ axis. This function is not singular at $z<1/B$. The branch, increasing with $r$, disappears at finite $r$ due to mutual interference of contributions with different $\beta$. As one can see, interference is an essential element of the phenomenon.
\subsection{Branch conversion}
Let us consider one of the functions $\psi_{I}$ (\ref{16}) which contains $f_{-}(\xi)[\varphi_{-}(\eta)+\varphi_{+}(\eta)]$. The sum of two functions has no logarithmic singularity at small $\eta$. In terms of the notation (\ref{18}), this function is $\psi_{1}+\psi_{2}$. As follows from Sec.~\ref{distr}, the two branches, $\psi_{1}$ and  $\psi_{2}$, exist at small $r$ and $1/2<z$.
More precisely, they exist to the right of the certain curve $R$ in Fig.~\ref{fig2} where two branches, $\psi_{2}$ and $\psi_{3}$, merge due to the condition $\partial\sigma_{2,3}/\partial\xi=0$.

The condition of the curve $R$, $\xi=1+\beta$, and equations of the type (\ref{20}) but for $i=2,3$ yield $\beta=\eta/\sinh^{2}\alpha$. As a result, the curve $R$ in Fig.~\ref{fig2} is determined by the form
$\xi_{R}(\eta)=1+\eta/\sinh^{2}\alpha$ or
\begin{equation}
\label{36aa}
r_{R}(z)=\frac{\sinh\alpha}{\sinh^{2}\alpha-1}\sqrt{(1-2z)(\sinh^{2}\alpha-2z)}\,.
\end{equation}
We consider $1<\sinh\alpha$. Close to the curve $R$, tangent derivatives are $\partial\sigma_{2,3}/\partial\eta =1/\tanh\alpha$. The derivative $\partial\sigma/\partial\xi\sim (\xi-\xi_{R})$ corresponds to $\psi_{2}$ to the right of the curve $R$ ($\xi_{R}<\xi$) and to $\psi_{3}$ to the left of the curve $R$ ($\xi<\xi_{R}$).

In contrast to $\psi_{1}$, the branch $\psi_{2}$ disappears to the left from $R$ since at this region there is no saddle point for it in the $\beta$-integration. In other words, the equation (\ref{20}) for $\psi_{2}$  does not have a solution at that region. This holds in the semiclassical limit only, that is not too close to the $z$ axis. To the left from the curve $R$ the branch $\psi_{2}$ exists close to the $z$ axis only where it serves to compensate the logarithmic singularity of $\psi_{1}$. So this is a realization of the state localized in the vicinity of the $z$ axis mentioned in Sec.~\ref{intr}.
\subsection{Crossing of the plane $\{z=0\}$ by the flux $\psi_{I}$}
As follows from Eq.~(\ref{20}), on the plane $\{z=0\}$, that is at $\xi=\eta=r$, the parameter $\beta$ satisfies the equation
\begin{equation}
\label{36a}
r\left[1+2\sqrt{\beta(1+\beta)}\sinh\alpha\right]=-\beta(1+\beta)\cosh^{2}\alpha
\end{equation}
and, according to Eqs.~(\ref{21a}) - (\ref{23}),
\begin{eqnarray}
\nonumber
&&\frac{\partial\sigma_{1}}{\partial z}=-\tanh\alpha-\frac{i}{\sqrt{r}},\hspace{0.3cm}z=0,\,\,r\ll 1\\
\label{36b}
&&\frac{\partial\sigma_{1}}{\partial r}=-\tanh\alpha+\frac{i}{\sqrt{r}}.
\end{eqnarray}
We see that the wave function $\psi_{I}$ increases with $z$ near the line $z=0$. Close to the left cusp, $\{z=-z_{s}, r=0\}$, the parameter $\beta=-1$ and, as follows from the relation (\ref{19}) for $\sigma_{1}$,
\begin{equation}
\label{36c}
|\psi_{I}(z=-z_{s},r=0)|\sim |\psi_{I}(z=z_{s},r=0)|\exp\left(-\frac{\pi B}{2}\right).
\end{equation}
The exponential comes from the $\eta$-integration from zero to one.

The caustics $A_{1}$ and $B_{3}$ are connected by classical paths, as shown in Fig.~\ref{fig2}. For this reason $|\psi_{I}|$ is the same (with the exponential accuracy) on these caustics. The
analogous statement is valid for caustics $A_{3}$ and $B_{1}$, indicated in Fig.~\ref{fig2} by the dashed curves, where $|\psi_{I}|$ is exponentially small according to Eq.~(\ref{36c}). So
$|\psi_{I}|$ exponentially decays at negative $z$.
\begin{figure}
\includegraphics[width=8.3cm]{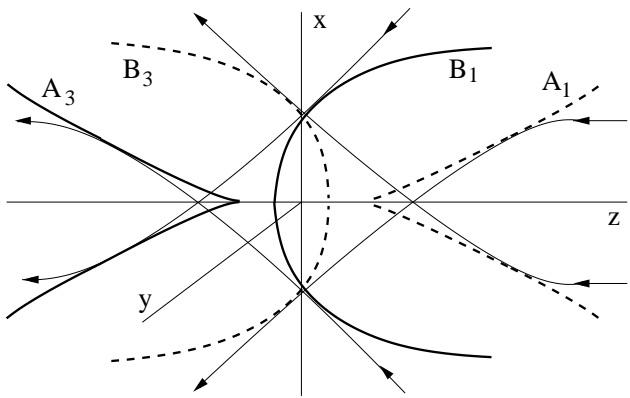}
\caption{\label{fig2a}Distribution of the particle flux associated with the wave function $\psi_{II}$. On the dashed caustics $\psi_{II}$ is exponentially small.}
\end{figure}
\section{DISTRIBUTION OF THE FLUX $\psi_{II}$}
\label{distrB}
As follows from Eqs.~(\ref{36}) and (\ref{36b}), the classical action becomes singular close to the center of the Coulomb field. At this region the semiclassical approximation is not true
and the branch $\psi_{I}$ is hybridized with $\psi_{II}$. The branch $\psi_{II}$ is determined by Eq.~(\ref{16a}).

We do not analyze here all details of $\psi_{II}$ which is similar to $\psi_{I}$. The caustics for the function $\psi_{II}$ are shown in Fig.~\ref{fig2a} where ones with exponentially small $\psi_{II}$ are related to dashed curves. The caustics are of the same form as in Fig.~\ref{fig2}. The analogue of the curve $R$ of Fig.~\ref{fig2} exists in Fig.~\ref{fig2a} in a symmetric way at negative $z$ (not shown in Fig.~\ref{fig2a}). For the branch $1b$ of $\psi_{I}$ at $z_{s}<z$ currents $j_{z}$ and $j_{r}$ are negative. For the analogous branch of $\psi_{II}$ at $z<-z_{s}$ the current $j_{z}$ is also negative but $j_{r}$ is positive. This is indicated in Figs.~\ref{fig2} and \ref{fig2a}. For the both cases the function $\beta(\xi,\eta)$ is the same. The relation, analogous to Eq.~(\ref{36c}), for the branch $\psi_{II}$ contains opposite sign of the exponential. So the branch $\psi_{II}$ exponentially decreases at positive $z$.
\section{HOW THE PARTICLE APPROACHES THE COULOMB CENTER}
\label{approach}
The flux on the Coulomb center comes from the right ($\psi_{I}$) and goes to the left ($\psi_{II}$). The functions $\psi_{I}$ and $\psi_{II}$ are hybridized in a vicinity of the Coulomb center. The total wave function (\ref{16b}) is a constant (with the exponential accuracy) on all caustics in Figs.~\ref{fig2} and \ref{fig2a}. The curves of classical velocities in these figures are symmetric with respect to the $x$ axis. The wave function exponentially decays inside shadow sides of caustics. The shadow side of the caustics $A_1$ and $B_1$ is between them. The same is valid for the caustics $A_3$ and $B_3$. The branches $\sigma_{1,2,3}$ describe the whole flux distribution. The branch $\sigma_4$ does not contribute to the semiclassical wave function since a proper saddle point is absent. In other words, all classical paths, described by $\sigma_4$, interfere down to zero not very close to the $z$ axis.

Since we are interested in the region close to the Coulomb center details of the flux, reflected from the caustics $B_1$ and $B_3$, are not important for this purpose. Therefore we focus on $\sigma_{1}$ at small $r$.
\begin{figure}
\includegraphics[width=8.0cm]{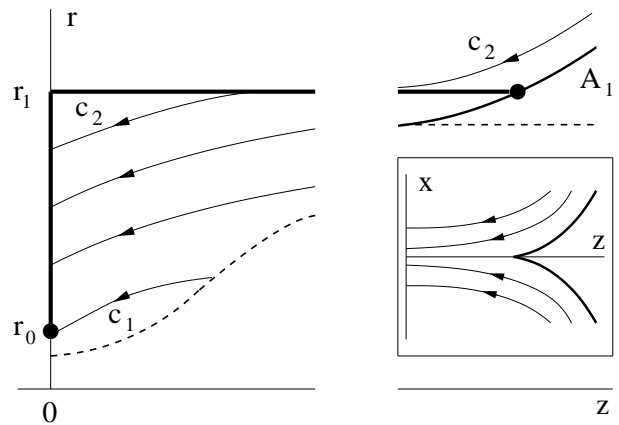}
\caption{\label{fig3}Curves $c_1$, $c_2$, and others of constant $|\psi|$ are also directions of momenta. The semiclassical approximation is not valid below the dashed curve described in the text. The left part of the interrupted plot is related to small $z$. Inset shows in the $\{x,z\}$ plane how the flux gets away from the cusp and moves to the Coulomb center (cusp driven tunneling).}
\end{figure}

At $r\sim z\ll 1$ the equation (\ref{20}) can be easily solved since in this case the parameter $1+\beta$ is small. The result is
\begin{eqnarray}
\nonumber
&&\frac{\partial\sigma_{1}}{\partial z}=-\frac{2\xi}{\xi+\eta}\tanh\alpha-\frac{2i\sqrt{\eta}}{\xi+\eta}\,,\hspace{0.2cm}r\sim z\ll 1\\
\label{38}
&& \frac{\partial\sigma_{1}}{\partial r}=-\frac{2\sqrt{\xi\eta}}{\xi+\eta}\tanh\alpha+\frac{2i\sqrt{\xi}}{\xi+\eta}\,,
\end{eqnarray}
which also corresponds to decrease of the wave function inside the shadow region. The continuity equation ${\rm div}\vec j=0$ in semiclassical approximation is equivalent to \cite{LANDAU1}
\begin{equation}
\label{38a}
\left(\nabla{\rm Im}\sigma\right)\left(\nabla{\rm Re}\sigma\right)=0,
\end{equation}
that is the momentum $\nabla{\rm Re}\sigma$ is directed along a curve of constant $|\psi|$ as in Fig.~\ref{fig3}. One can check that expressions (\ref{37}) and  (\ref{38}) satisfy the 
condition (\ref{38a}).

In Fig.~\ref{fig3} the local angle $\chi$ between the tangent direction of the curves $c_{1,2}$ and the $z$ axis is given by the relation
\begin{equation}
\label{38b}
\frac{\partial{\rm Im}\sigma}{\partial r}\tan\chi=-\frac{\partial{\rm Im}\sigma}{\partial z}.
\end{equation}
As follows from Eqs.~(\ref{37}) and (\ref{38}), at $r\ll 1$
\begin{eqnarray}
\nonumber
&&\tan\chi=\frac{rz_s}{2z(z_s-z)}\,,\hspace{0.2cm}r\ll z<z_s\\
\label{38c}
&&\tan\chi=\left(\frac{\sqrt{r^2+z^2}-z}{\sqrt{r^2+z^2}+z}\right)^{1/2},\hspace{0.2cm}r\sim z\ll 1.
\end{eqnarray}
According to these equations, the curves $c_{1}$, $c_2$, and others in Fig.~\ref{fig3} are directed by $45^{\circ}$ with respect to the $r$ axis at $z=0$. At $z\sim 1$ these curves are almost parallel to the $z$ axis.

Another important aspect is a border of applicability of the semiclassical approximation used. This approximation holds when the term with quantum corrections (one with second derivatives in
Eq.~(\ref{6})) to the Hamilton-Jacobi equation is small. Substituting expressions (\ref{37}) and (\ref{38}) into Eq.~(\ref{6}), one can conclude that the semiclassical approximations is violated below the dashed curve in Fig.~\ref{fig3} which is $r\sim 1/B^2$ at $z<1/B^2$, $r\sim\sqrt{z}/B$ at $1/B^2<z<1/2$, and $r\sim 1/B$ at larger $z$. In physical units the scale $1/B$ corresponds to the de Broglie wave length defined in Sec.~\ref{gen}. The scale $1/B^2$ is Bohr radius (Sec.~\ref{gen}) which is the shortest spatial scale of the Coulomb problem. As it should be, the semiclassical approximation is not valid too close to the $z$ axis \cite{LANDAU1}.

It follows from Fig.~\ref{fig3} that the curve $c_2$ is entirely in the semiclassical region and approaches to the caustic $A_1$. But the curve $c_1$ enters under the dashed curve. The distribution of $|\psi|$ in Fig.~\ref{fig3} relates to the branch $\psi_1$.

The inset in Fig.~\ref{fig3} shows the flux flow in the shadow region of the caustic $A_1$. According to general properties of caustics, $|\psi_{1}|$ decays at the shadow side away from $A_1$. Arrowed curves close to the caustic $A_1$ in the inset in Fig.~\ref{fig3} relate to constant $|\psi_{1}|$. The closer the curve is to $A_1$ the larger $|\psi_{1}|$ is. The same is valid if to approach the $z$ axis at the region left from the cusp, where the modulus of the wave function reaches its maximal value on the $z$ axis. That value is kept along the whole segment of the $z$ axis, from the cusp to the center.

The caustic $A_1$ in Fig.~\ref{fig3} is associated with the branch $1$ of $\psi_{I}$. The analogous branch (related to $\psi_{II}$) is connected with the caustic $A_3$ and supplements Fig.~\ref{fig3} by the mirror reflection. In this way the flux is continued from positive to negative $z$. So the particle flux in Fig.~\ref{fig2}, from the caustic $A_1$ to $A_3$, is localized at the narrow channel around the $z$ axis. This can be called cusp driven tunneling.

The same type of channel, analogous to the wave $\psi_{1}$ to the left from the cusp, is associated with a cusp caustic in optics and acoustics. It coexists with the wave, analogous to $\psi_{2}$, which escapes from the cusp caustic piercing it. This results in interference oscillations, extended far to the left from the cusp, in the total light amplitude \cite{BERRY}. These oscillations are unavoidable since they result from topological properties, namely, dislocations in the spatial distribution of the total light amplitude \cite{BERRY}. Without the channel it would be solely the wave, analogous to $\psi_{2}$, which does not lead to the interference oscillations. At large distances from the cusp, the channel is smeared out in space due to non-semiclassical effects. In our case such distances are not involved since the center is close to the cusp.

The tunneling probability can be defined as the ratio of densities at the center and at large distances
\begin{equation}
\label{39}
w\simeq\Bigg|\frac{\psi(r_0,0)}{\psi(r_1,\infty)}\Bigg|^{2}.
\end{equation}
The parameters $r_0$ and $r_1$ are indicated in Fig.~\ref{fig3}. The point $\{r_0,0\}$ is not exactly at the center but when $r_0\sim 1/B^2$ the expression (\ref{39}) does not differ from the exact probability in the exponential approximation. The tunneling probability has the form
\begin{equation}
\label{40}
w\sim\exp\left\{-2B\,{\rm Im}\left[\sigma_{1}(r_0,0)-\sigma_{1}(r_1,\infty)\right]\right\}
\end{equation}
The infinity point can be substituted by one on the caustic $A_1$ as shown in Fig.~\ref{fig3} since to the right from the caustic $\sigma$ is real. 

If in the expression (\ref{40}) $1/B^2<r_0<r_1$ and $1/B<r_1$ then one can use the semiclassical approximation along the path (two thick lines) between the two dots in Fig.~\ref{fig3}. For convenience, we take in addition $r_1\ll 1$. Then the first equation (\ref{38}) simply gives
\begin{equation}
\label{40a}
{\rm Im}\left[\sigma_{1}(r_1,\infty)-\sigma_{1}(r_1,0)\right]=-2\sqrt{r_1}\,.
\end{equation}
The second equation (\ref{38}) produces
\begin{equation}
\label{40b}
{\rm Im}\left[\sigma_{1}(r_1,0)-\sigma_{1}(r_0,0)\right]=2\sqrt{r_1}-2\sqrt{r_0}\,.
\end{equation}
By means of Eqs.~(\ref{40}) - (\ref{40b}) the ratio of densities at points $\{r_0,0\}$ and $\{r_1,\infty\}$ (\ref{39}) can be written in the form
\begin{equation}
\label{40c}
w\sim\exp\left(-4B\sqrt{r_0}\right).
\end{equation}

Equation (\ref{40c}) holds when $1/B^2<r_0$ and formally that expression is exponentially small. It is clear that one can continue $r_0$ down to the border of applicability of the result (\ref{40c}), namely to put $r_0\sim 1/B^2$. One can conclude from here that tunneling probability is not exponentially small.
\begin{figure}
\includegraphics[width=6.5cm]{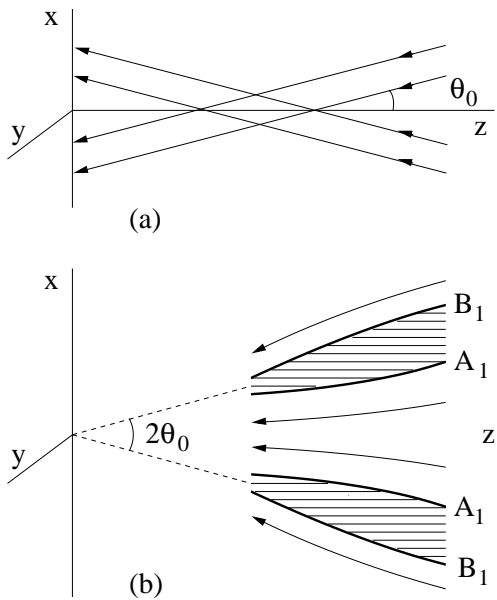}
\caption{\label{fig4}(a) Conical particle flux from the right in the absence of the Coulomb potential. (b) Split of the conical flux by the Coulomb field at large $z$. Trajectories are reflected from caustics which are continuations of ones in Fig.~\ref{fig2}. The dashed parts are caustic shadows.}
\end{figure}
\section{INCIDENT FLUX FAR FROM THE COULOMB CENTER}
\label{far}
In the absence of the Coulomb potential the flux from the left to the right relates to $\sigma=-2z$ and consequently to the plane wave $\psi=\exp(-2iBz)$ in the whole space. If we add the Coulomb potential that plane wave is hardly influenced far to the right but becomes strongly violated at finite distances. In particular, the caustic is formed as shown in Fig.~\ref{fig1}.
This situation corresponds to $\beta=-i/B$ as mentioned in Sec.~\ref{scat}.

Suppose again that the Coulomb potential is absent but the incident flux from the right is not a plane wave but one related to $\sigma=-z\tanh\alpha\pm r/\cosh\alpha$. The corresponding exact wave function is expressed through the Bessel function
\begin{equation}
\label{41}
\psi=\exp\left(-2iBz\tanh\alpha\right)J_{0}\left(\frac{2Br}{\cosh\alpha}\right).
\end{equation}
In this case the velocity distribution is shown in Fig.~\ref{fig4}(a).

If we add the Coulomb potential the velocity field becomes strongly deformed. The new velocity distribution is illustrated in Fig.~\ref{fig4}(b) where
\begin{equation}
\label{41a}
\cot\theta_0=\sinh\alpha.
\end{equation}
It is easy to qualitatively understand the features of that distribution. The flux, locally parallel as in Fig.~\ref{fig4}(a), is reflected by the Coulomb force upward when the polar angle $\theta$ ($\tan\theta=r/z$) exceeds $\theta_0$. This flux is reflected downward when $\theta$ is less than $\theta_0$. There is the shadow region between these limits as in Fig.~\ref{fig4}(b).

If $n$ is a direction normal to a caustic surface then the wave function decays inside the shadow region as
\begin{equation}
\label{42}
\psi\sim\exp\left[-\left(n/l\right)^{3/2}\right],
\end{equation}
where $l\sim(\lambda^{2}z)^{1/3}$ in physical units. Here $\lambda$ is the wave length defined in Sec.~\ref{gen}. Eq (\ref{42}) is a usual form for caustics \cite{LANDAU3}. As follows from Eqs.~(\ref{26}) and (\ref{28}), the distance between caustics $A_1$ and $B_1$ at large $z$ is proportional, in physical units, to $\sqrt{ze^{2}/E}$ which is larger than $l$. So there is a real shadow between the caustics.

We see that properties of the incident flux at large distances strongly determine the wave function at the Coulomb center.
\section{DISCUSSIONS}
\label{disc}
Probability of tunneling across a one-dimensional static potential barrier is exponentially small when the barrier is almost classic. The wave function decays inside the barrier since in the classically forbidden region the wave vector is imaginary.

Below some not rigorous arguments are given. When dimensionality of a problem is higher than one the situation can be more complicated since the wave vector has more than one components in space. The sum of squared components, which is a kinetic energy, is negative under the barrier. In the classical manner this can be written in the form (compare with Eq.~(\ref{10a}))
\begin{equation}
\label{43}
k^{2}_{r}+k^{2}_{z}+\frac{1}{2\sqrt{r^2+z^2}}=1.
\end{equation}
In our case the $z$ axis is the tunneling direction and tunneling probability is determined by ${\rm Im}\,k_z$. The smaller is this value the larger is the probability. Eq.~(\ref{43})
formally says that a reduction of ${\rm Im}\,k_z$ is connected with increasing of ${\rm Im}\,k_r$.

A localization near the $z$ axis is equivalent to a large ${\rm Im}\,k_r$ at that region and therefore leads to increase of tunneling rate along the $z$ direction. A localization of the density around the $z$ axis is possible but, as a ``payment`` for this, the wave function has a non-physical singularity on that line. Therefore one has to add the second branch to compensate the singularity. But the second branch exponentially increases away from the $z$ axis which violates, at the first sight, the construction.

The above statement holds when some particular value of $k_{r}$ is chosen. If we consider a continuous set of waves with various $k_{r}$, properties of the resulting wave function can be completely different. It is exactly our case. Due to an interference in the set of second branches, they compensate each other at finite $r$ excepting a narrow (non-semiclassical) vicinity of the $z$ axis. This occurs since for the second branch the $\beta$ integration is entirely on the same descent side of the saddle. As a result, only the state decaying from the $z$ axis (channel) survives. An analogous channel is associated with a cusp caustic in optics and acoustics due to the similar phenomenon of getting away from a caustic cusp.

To get a cusp caustic one should have a convergent incident wave at large distances. In our case the cusp position on the $z$ axis, $e^2/(E\sin^2\theta_{0})$, is determined by the characteristic angle $\theta_{0}$ of the incident conical flux. When the incident flux is just a plane wave, as in the conventional scattering problem, there is no cusp caustic. In this case only the usual one exist as in Fig.~\ref{fig1}.

An origin of the cusp is based on global properties of the flux. The Coulomb field separates the incident conical flux, with the characteristic angle $\theta_{0}$ in Fig.~\ref{fig4}(a), according to geometrical rules. Namely, all rays with larger angles are turned upward and ones with smaller angles are turned downward. This is equivalent to a separation of the two streams by caustics. In continuation toward the Coulomb center, the internal caustic shrinks to a cusp point on the $z$ axis before the center. Otherwise the flux classically reaches the Coulomb center. The flux gets away from the cusp and moves to the center along the above channel containing the $z$ axis. Not too close to the center the radius of the channel is proportional to de Broglie wave length.

This is cusp driven tunneling when the cusp forms the flux on the Coulomb center. In the vicinity of the center the flux passes the bottle neck of the size of the nuclear Bohr radius. Then at negative $z$ the flux moves to the left.

The fusion probability is proportional to a neutron yield. The neutron yield of the process depends on the weight function $c_{I}(\beta)$ (\ref{16}) and, therefore, on details of the incident flux. With the choice (\ref{17}) the yield is not exponentially small as $\exp(-6174.9)$ (see Sec.~\ref{gen}) but it can be ``normally'' small due to a preexponential factor in the tunneling probability. One can increase this factor by variation of a pre-factor in the weight function (\ref{17}).

We consider in the text the axially symmetric wave function of two deuterons related to the magnetic quantum number $m=0$. A finite $m$ reduces the effect due to formation of a centrifugal repulsive barrier.

In this paper we briefly mention experimental schemes for formation of a particle flux resulting in the cusp phenomenon.

One of experimental ways to produce the conical flux of deuterons (\ref{41}), shown in Fig.~\ref{4}(a), is to confine them in a long tube with rigid walls, for example, in a nanotube. Another way is to push deuterons (atoms) to pass through a diffraction grid of a conical shape. Since the de Broglie wave length is of the order of $1~\AA$, one can use a natural crystal lattice. A setup with slits also can be used. This is a situation of quantum lens. We will discuss the details elsewhere.

Above two bare deuterons are considered. From the practical standpoint it can be more convenient to deal with a substance (heavy water, for example) consisted of molecules with deuterium. Inside a single molecule of heavy water the deuterons are in the well with vibration energy levels. An external laser radiation can influence a quantum state of deuterons in the well. One can put a question whether the radiation is able to create something like cusp state in the well and which pulse shape should be used for this purpose.

Anyway a formation of a required particle flux in experiments needs a detailed study which is outside this paper.
\section{CONCLUSIONS}
Fusion of two bare (supposed to be isolated from everything) deuterons of room temperature energy can be possible. The penetration across the Coulomb barrier, which is the main obstacle for low-energy fusion, strongly depends on a form of the incident flux on the center at large distances from it. In contrast to the usual scattering, the incident wave is not a single plane wave but a superposition of plane waves of the same energy and various directions, for example, a convergent conical flux. As a result of interference, the wave function close to the Coulomb center is determined by the cusp caustic which is probed by de Broglie waves. The particle flux gets away from the caustic cusp and moves to the Coulomb center providing the not small probability of fusion (cusp driven tunneling).
\acknowledgments
The author is grateful to P. Cooper, J. Engelfried, and M. Kirchbach for discussions of related topics.

\end{document}